\documentclass{Interspeech}

\interspeechcameraready

\usepackage{graphicx}
\usepackage{url}
\usepackage{amsmath}
\usepackage{amssymb}
\usepackage{graphicx}
\usepackage{subcaption}
\usepackage{booktabs} 

\title{U-SAM: An Audio Language Model for Unified Speech, Audio, and Music Understanding}

\author[affiliation={1}]{Ziqian}{Wang}
\author[affiliation={2}]{Xianjun}{Xia}
\author[affiliation={1}]{Xinfa}{Zhu}
\author[affiliation={1}\ast]{Lei}{Xie}

\affiliation{Audio, Speech and Language Processing Group (ASLP@NPU), School of Software}{Northwestern Polytechnical University, Xi'an}{China}
\affiliation{}{Bytedance Inc., Sydney}{Australia}

\email{zq\_wang@mail.nwpu.edu.cn, lxie@nwpu.edu.cn}

\usepackage{comment}

\begin{document}

\maketitle

\begin{abstract}
    
The text generation paradigm for audio tasks has opened new possibilities for unified audio understanding. However, existing models face significant challenges in achieving a comprehensive understanding across diverse audio types, such as speech, general audio events, and music. Furthermore, their exclusive reliance on cross-entropy loss for alignment often falls short, as it treats all tokens equally and fails to account for redundant audio features, leading to weaker cross-modal alignment. To deal with the above challenges, this paper introduces U-SAM, an advanced audio language model that integrates specialized encoders for speech, audio, and music with a pre-trained large language model (LLM). U-SAM employs a Mixture of Experts (MoE) projector for task-aware feature fusion, dynamically routing and integrating the domain-specific encoder outputs. Additionally, U-SAM incorporates a Semantic-Aware Contrastive Loss Module, which explicitly identifies redundant audio features under language supervision and rectifies their semantic and spectral representations to enhance cross-modal alignment. Extensive experiments demonstrate that U-SAM consistently outperforms both specialized models and existing audio language models across multiple benchmarks. Moreover, it exhibits emergent capabilities on unseen tasks, showcasing its generalization potential. Code is available\footnote{\url{https://github.com/Honee-W/U-SAM/}}.

\end{abstract}

\renewcommand{\thefootnote}{\fnsymbol{footnote}}  
\footnotetext{Corresponding author.}

\section{Introduction}

\begin{figure*}[]
  \centering
  \includegraphics[width=1.0\linewidth]{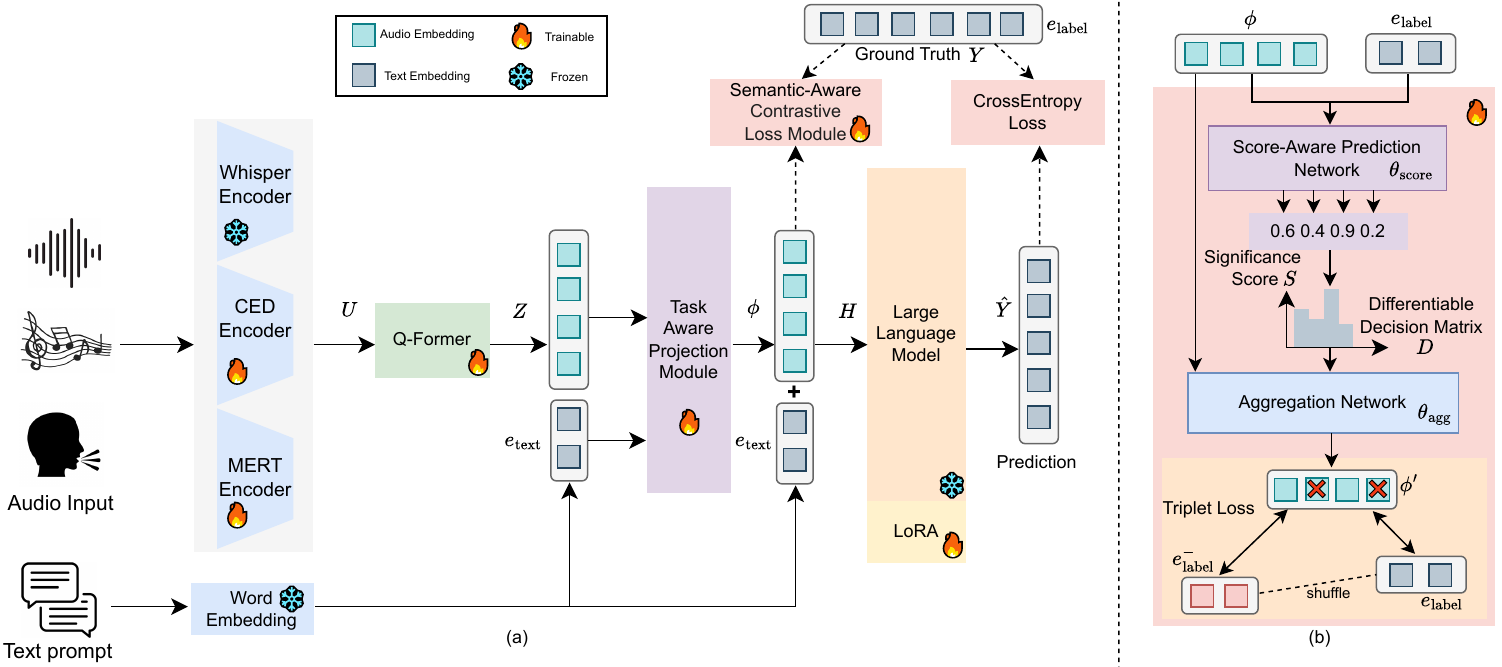}
\caption{(a) U-SAM architecture: Integrates three encoders (Whisper, CED, and MERT) to extract multi-scale audio features, refined by a Q-Former and Task-Aware Projection Module (TAPM) for task-specific adaptation. The adapted features, combined with text prompts, are processed by a large language model (LLM) for semantic understanding and response generation. (b) Semantic-Aware Contrastive Loss Module (SACLM): Predicts significance scores for audio embeddings via the Score-Aware Prediction Network, generates a decision matrix for frame selection, refines frames with an Aggregation Network, and aligns audio embeddings with text using a Triplet Loss.}
    \label{fig:pipeline}
\end{figure*}

\label{sec:intro}
Large language models (LLMs) have achieved remarkable success in natural language processing (NLP), demonstrating impressive capabilities across a wide range of tasks~\cite{zhao2023survey}. Their ability to reason and generate coherent outputs stems from extensive pretraining on large-scale text data. Recently, research has expanded beyond text to integrate LLMs with other modalities such as images, video, and audio. While vision language models (VLMs)~\cite{alayrac2022flamingo, dai2023instructblipgeneralpurposevisionlanguagemodels, li2023blip} have been widely studied and show advanced multimodal capabilities, audio language models (ALMs) are still an emerging area of research. Beyond visual and language perception, understanding audio is essential for creating intelligent systems that can effectively interact with real-world environments.

Audio language models (ALMs) can be broadly categorized into two approaches. The first approach, exemplified by CLAP~\cite{elizalde2023clap}, uses an audio encoder and a text encoder with contrastive learning to align audio and text representations~\cite{deshmukh2022audio, xin2023improving, yuan2024tclaptemporalenhancedcontrastivelanguageaudio}. These models enhance audio understanding by leveraging natural language supervision but are limited by the scope of their paired data and the reliance on fixed alignment objectives.  The second approach adopts an audio encoder combined with a large language model to consider audio tasks as open-ended text generation tasks. Representative models include Qwen-Audio~\cite{chu2023qwenaudioadvancinguniversalaudio} and Qwen2-Audio~\cite{chu2024qwen2}, which utilize the Whisper encoder~\cite{radford2022robustspeechrecognitionlargescale}, Pengi~\cite{deshmukh2023pengi} based on CLAP’s audio transformer backbone, and LTU~\cite{gong2023listen} and GAMA~\cite{ghosh2024gama}, which employ AST encoder~\cite{gong2021ast}. While effective for their intended tasks, these single-encoder models are often specialized for specific domains, such as ASR or sound event detection, and struggle to generalize across diverse audio types. Recent works like SALMONN~\cite{tang2023salmonn} incorporate multiple audio encoders, such as Whisper and BEATs~\cite{chen2022beats}, to capture complementary information for speech, music, and environmental sounds. However, these methods rely on simple concatenation for feature fusion, treating all encoder outputs equally without considering task-specific relevance, which limits their ability to adaptively integrate features from diverse audio inputs, leaving room for improvement in task-aware audio understanding~\cite{wang2024speechlanguagemodelsfail}.

Beyond architectural limitations, existing ALMs often rely on cross-entropy loss for audio-text alignment, which struggles with the inherent differences between audio and text modalities. Audio sequences are significantly longer than text and contain repetitive or irrelevant information, complicating alignment. By treating all tokens equally, cross-entropy loss fails to distinguish meaningful features from redundant ones, weakening the model’s ability to establish effective audio-text associations~\cite{reimers2019sentence, Fu_2024_CVPR}.

To address these limitations, we propose \textbf{U-SAM}, a \textbf{U}nified audio language model that integrates specialized audio encoders for \textbf{S}peech, \textbf{A}udio, and \textbf{M}usic with a pre-trained large language model (LLM). The key contributions of U-SAM are as follows: (1) a Task-Aware Projection Module (TAPM) that dynamically routes and fuses embeddings from multiple audio encoders according to the corresponding task, enabling task-aware feature fusion and deeper multimodal alignment; (2) a Semantic-Aware Contrastive Loss Module (SACLM) that refines redundant audio frames under language supervision, achieving more precise and consistent audio-text alignment; and (3) a generalized framework that considers audio tasks as text generation, allowing U-SAM to handle a wide range of tasks while demonstrating emergent capabilities on unseen tasks. Extensive experiments demonstrate that U-SAM outperforms existing methods across multiple benchmarks, underscoring its potential to advance comprehensive and generalized audio understanding.


\section{Methodology}
\label{sec:method}
In this section, we present the U-SAM architecture in Section \ref{model}, detailing its components, including specialized audio encoders, the Task-Aware Projection Module, and integration with a pre-trained LLM. The training objectives, supervised cross-entropy loss, and the proposed Semantic-Aware Contrastive Loss are discussed in Section \ref{loss}.

\subsection{Model Architecture}
\label{model}
The U-SAM architecture, shown in Fig. \ref{fig:pipeline}(a), integrates three specialized audio encoders and a Q-Former with TAPM to fuse complementary audio embeddings for downstream tasks. The fused embeddings, combined with text prompts, are fed into a pre-trained LLM to generate task-specific responses.

\begin{figure}[]
  \centering
  \includegraphics[width=0.9\linewidth]{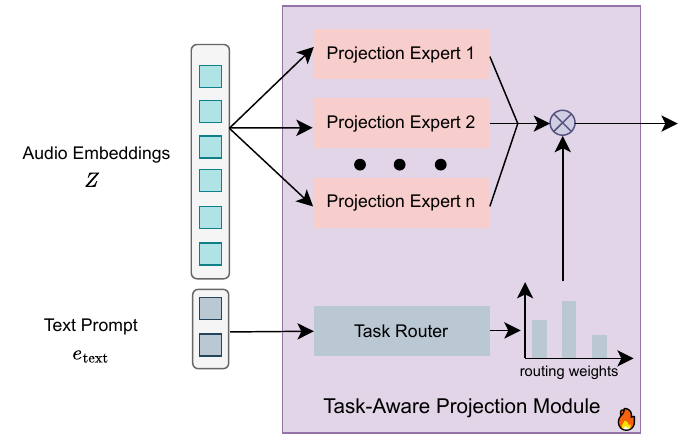}
\caption{Task-Aware Projection Module (TAPM). TAPM uses a Mixture of Experts to dynamically project audio embeddings based on routing weights derived from text prompt embeddings. The weighted outputs of $n$ experts are combined to generate task-adaptive features for the downstream LLM.}
\label{fig:moe}
\end{figure}

\subsubsection{Specialized Audio Encoders}
U-SAM employs three specialized encoders to capture the diverse characteristics of speech, audio, and music, enabling comprehensive audio understanding. The Whisper encoder~\cite{radford2022robustspeechrecognitionlargescale}, pre-trained on large-scale weakly supervised data, excels in speech-related tasks such as automatic speech recognition (ASR) and automatic speech translation (AST) by extracting features like phonetic structure and prosody. The CED encoder~\cite{dinkel2024ced} focuses on non-speech audio, modeling broad spectral and temporal patterns through self-supervised learning, making it effective for tasks like audio event classification. The MERT encoder~\cite{yizhi2023mert} specializes in music, capturing harmonic, rhythmic, and melodic structures, which are critical for tasks such as music captioning.
These encoders output sequences of varying lengths, which are padded and concatenated along the channel dimension to form a unified input:
\begin{equation}
\label{eq:encoder}
U = \text{Concat}((E_{\text{Whisper}}(X)), (E_{\text{CED}}(X)), (E_{\text{MERT}}(X))),
\end{equation}
where \( X \) is the input audio sequence, and \( E_{\text{Whisper}} \), \( E_{\text{CED}} \), and \( E_{\text{MERT}} \) represent the Whisper, CED, and MERT, respectively.

\subsubsection{Window-level Q-Former}
U-SAM employs a Window-level Q-Former~\cite{li2023blip, tang2023salmonn} which allows the sequence length of the output embeddings to adapt to the input audio, preserving rich temporal information. Given audio embeddings \( U \in \mathbb{R}^{B \times T \times D} \) and trainable queries \( Q \in \mathbb{R}^{N \times D} \), the Q-Former computes query-aware representations \( Z \) through self-attention and cross-attention mechanisms:  
\begin{equation}
Z = \text{CrossAttention}(\text{SelfAttention}(Q), U),
\end{equation}  
where \( N \) is the number of queries, \( T \) is the variable input sequence length, and \( D \) is the feature dimension.

\subsubsection{Task-Aware Projection Module (TAPM)}
TAPM selects and integrates task-specific audio embeddings based on the input text prompt, as shown in Fig. \ref{fig:moe}. It utilizes a Mixture of Experts (MoE) approach~\cite{cai2024surveymixtureexperts}, where each expert specializes in a different subset of tasks. The routing weights for expert selection are computed from the text prompt embedding:
\begin{equation}
w = \text{softmax}(W_r e_{\text{text}}),
\end{equation}
where \( W_r \) is a learnable weight matrix and \( e_{\text{text}} \) is the text prompt embedding. The final task-adaptive features are computed by combining the outputs of all experts:
\begin{equation}
\phi  = \sum_{i=1}^n w_i \cdot \text{Expert}_i(Z),
\end{equation}
where \( \text{Expert}_i(\cdot) \) processes the audio embeddings \( Z \) and \( w_i \) are the routing weights.

\subsubsection{Integration with LLM via LoRA}
To efficiently fine-tune the LLM for multimodal tasks, U-SAM uses LoRA (Low-Rank Adaptation)~\cite{hu2021lora}, which updates only the query and value weight matrices in the self-attention layers while keeping other parameters frozen. This significantly reduces the computational overhead.

The LoRA update can be expressed as:
\begin{equation}
W = W_{\text{base}} + \Delta W, \quad \Delta W = A \cdot B,
\end{equation}
where \( W_{\text{base}} \) is the pre-trained weight matrix, and \( A \in \mathbb{R}^{d \times r} \), \( B \in \mathbb{R}^{r \times d} \) are low-rank trainable matrices with \( r \ll d \).

During training, only \( A \) and \( B \) are updated. With input embeddings \( H \), formed by concatenating \( \phi \) and \( e_{\text{text}} \) along the channel dimension, the adapted attention mechanism is:
\begin{equation}
\text{Attention}(H) = \text{softmax}\left(\frac{H W_q H^T}{\sqrt{d}}\right) H W_v,
\end{equation}
where \( W_q \) and \( W_v \) include the low-rank updates from LoRA.

\subsection{Training Objectives}
\label{loss}

\subsubsection{Supervised Cross-Entropy Loss}
The supervised cross-entropy loss is used to align predicted outputs with target text tokens. Given the input embeddings \( H  \) and the corresponding ground-truth text \( Y \), the model predicts token logits \( \hat{Y} \):
\begin{equation}
\hat{Y} = \theta_{\text{LLM}}(H),
\end{equation}
where \( \theta_{\text{LLM}} \) represents the LLM. The cross-entropy loss is computed as:
\begin{equation}
\mathcal{L}_{\text{CE}} = -\sum_{i=1}^N y_i \log(\hat{y}_i),
\end{equation}
where \( y_i \) and \( \hat{y}_i \) are the ground truth and predicted token probabilities, respectively, and \( N \) is the sequence length.

\subsubsection{Semantic-Aware Contrastive Loss}
To enhance cross-modal alignment between audio and text embeddings, we propose a Semantic-Aware Contrastive Loss, which uses a contrastive triplet loss to identify key audio features and align them with text semantics, as illustrated in Fig.\ref{fig:pipeline}(b).

Given audio embeddings \( \phi \in \mathbb{R}^{B \times T_a \times D} \) and text embeddings \( T \in \mathbb{R}^{B \times T_t \times D} \), the temporal dimensions of both are aligned through interpolation. A significance score \( S \in [0, 1]^{B \times T_a} \) is predicted for each audio frame based on the similarity between the audio and text embeddings:
\begin{equation}
S = \theta_{\text{score}}(\phi, T),
\end{equation}
where \( \theta_{\text{score}} \) computes the relevance of each audio frame to the text.

A decision matrix \( D \) is generated by thresholding \( S \) at 0.5, selecting the most significant audio frames. The aggregation of these frames yields refined audio embeddings \( \phi' \):
\begin{equation}
\phi' = \theta_{\text{agg}}(\phi, D),
\end{equation}
where \( \theta_{\text{agg}} \) is the aggregation function that combines the selected frames into a more compact and semantically relevant representation.

To enforce alignment, a contrastive triplet loss~\cite{Dong_2018_ECCV} is applied between the aggregated audio embeddings \( \phi' \), the corresponding text embeddings \( T \), and randomly sampled negative text embeddings \( T^- \). The triplet loss is defined as:
\begin{equation}
\mathcal{L}_{\text{triplet}} = \max \left( d(\phi', T) - d(\phi', T^-) + \alpha, 0 \right),
\end{equation}
where \( d(\cdot, \cdot) \) is the cosine distance function and \( \alpha \) is a margin hyperparameter.

To encourage sparsity in the significance scores \( S \), we add a regularization term:
\begin{equation}
\mathcal{L}_{\text{SAC}} = \mathcal{L}_{\text{triplet}} + \lambda \lVert S \rVert_1,
\end{equation}
where \( \lVert S \rVert_1 \) is the \( L_1 \)-norm of the significance scores \( S \), and \( \lambda \) is a regularization parameter, set to 0.01, to control the sparsity of the selected audio frames.

The final loss is defined as \( \mathcal{L} = \alpha \mathcal{L}_{\text{CE}} + (1-\alpha) \mathcal{L}_{\text{SAC}} \), with \( \alpha \) fixed at 0.5.

\section{Experiments}
\label{sec:exp}

\subsection{Data Specifications}
U-SAM is trained and evaluated on diverse datasets across multiple tasks. For automatic speech recognition (ASR), it uses LibriSpeech~\cite{panayotov2015librispeech} and evaluates on the test-clean/-other subsets with word error rate (WER). Speech-to-text translation (En2Zh) employs CoVoST2-En2Zh~\cite{wang2020covost}, evaluated by BLEU4. audio captioning (AAC) combines AudioCaps~\cite{kim2019audiocaps}, Clotho~\cite{drossos2020clotho}, and WavCaps~\cite{mei2024wavcaps} for training and evaluates on AudioCaps using METEOR and SPIDEr. Music captioning (MC) uses MusicCaps~\cite{agostinelli2023musiclm}, evaluated with BLEU4 and RougeL. Details are summarized in Table~\ref{tab:training_evaluation}.

\subsection{Model Specifications}

U-SAM integrates Whisper-Large v2\footnote{\url{https://huggingface.co/openai/whisper-large-v2}} for speech, CED\footnote{\url{https://huggingface.co/mispeech/ced-base}} for general audio, and MERT\footnote{\url{https://huggingface.co/m-a-p/MERT-v1-330M}} for music. The backbone LLaMA 3.1\footnote{\url{https://huggingface.co/meta-llama/Llama-3.1-8B}} retains frozen weights, with LoRA modules (rank 8) applied for lightweight adaptation. The Q-Former uses a single trainable query. TAPM includes three MLP experts and an MLP router, while SACLM employs MLPs for both scoring and aggregation. Only the Q-Former, TAPM, SACLM, and LoRA are trainable, totaling 41M parameters (0.39\%). Training uses a batch size of 64, AdamW~\cite{loshchilov2019decoupledweightdecayregularization} with a 5e-5 learning rate, 1e-6 weight decay, linear decay schedule, 0.13 warm-up ratio, and runs on 4×NVIDIA H20 GPUs for 5 days.

\begin{table}[t]
\centering
\caption{Datasets and evaluation metrics used for U-SAM.}
\label{tab:training_evaluation}
\resizebox{0.5\textwidth}{!}{
\begin{tabular}{@{}lcc@{}}
\toprule
\textbf{Task} & \textbf{Training/Test Data}              & \textbf{Eval Metrics} \\ \midrule
ASR           & LibriSpeech / LibriSpeech test-clean/-other & \%WER                 \\
En2Zh         & CoVoST2-En2Zh / CoVoST2-En2Zh             & BLEU4                 \\
AAC           & AudioCaps + Clotho + WavCaps / AudioCaps            & METEOR \textbar{} SPIDEr \\
MC            & MusicCaps / MusicCaps                     & BLEU4 \textbar{} 
RougeL         \\ \bottomrule
\end{tabular}
}
\end{table}

\subsection{Baselines}
U-SAM is compared against i) state-of-the-art specialized models: Whisper for ASR, MT-Rev~\cite{wang2020covost} for En2Zh, HASAT-BART~\cite{mei2024wavcaps} for AAC, and LP-MusicCaps~\cite{doh2023lp} for MC; ii) general-purpose audio language models: LTU~\cite{gong2023listen}, SALMONN~\cite{tang2023salmonn}, Pengi~\cite{deshmukh2023pengi}, AudioGPT~\cite{huang2024audiogpt}, and GAMA~\cite{ghosh2024gama}. Original checkpoints released by the authors are used to obtain the aforementioned system performances.

\begin{table}[t]
\centering
\caption{Comparsion of U-SAM with baselines on evaluation datasets.}
\label{tab:experimental_results}
\resizebox{0.5\textwidth}{!}{
\begin{tabular}{@{}lcccc@{}}
\toprule
      & \textbf{ASR}$\downarrow$ & \textbf{En2Zh}$\uparrow$ & \textbf{AAC}$\uparrow$ & \textbf{MC}$\uparrow$ 
\\ 
\textbf{Models}                  & \%WER       & BLEU4         & METEOR\textbar{}SPIDEr       & BLEU4\textbar{}RougeL      
\\ \midrule
Whisper~\cite{radford2022robustspeechrecognitionlargescale}              & (2.2, 5.1)           & -                      & -                                     & -                                   \\
bilingual MT-Rev~\cite{wang2020covost} & -                   & \textbf{38.9}                   & -                                     & -                                   \\
HASAT-BART~\cite{mei2024wavcaps} & -                   & -                      & 25.0\textbar{}48.5                    & -                                   \\
LP-MusicCaps~\cite{doh2023lp}      & -                   & -                      & -                                     & \textbf{6.1}\textbar{}21.5                   \\ \midrule
LTU~\cite{gong2023listen}                  & -                    & -                      & 25.5\textbar{}45.1                    & -                                   \\
SALMONN~\cite{tang2023salmonn}              & (2.1, \textbf{4.9})           & 33.1                   & 24.0\textbar{}40.3                    & 5.5\textbar{}21.8                   \\
Pengi~\cite{deshmukh2023pengi}                & -                    & -                      & 24.3\textbar{}41.8                    & -                                   \\
AudioGPT~\cite{huang2024audiogpt}             & -                    & -                      & 22.5\textbar{}38.0                    & -                                   \\
GAMA~\cite{ghosh2024gama}                 & -                    & -                      & 26.1\textbar{}47.8                    & -                                   \\
\textbf{U-SAM (ours)} & (\textbf{2.0}, 5.0)  & 38.3          & \textbf{26.7\textbar{}49.0}           & 6.0\textbar{}\textbf{22.5}          \\ \bottomrule
\end{tabular}%
}
\end{table}

\subsection{Experimental Results and Discussion}

Table~\ref{tab:experimental_results} compares U-SAM with various baseline models across four evaluation tasks: ASR, AST-En2Zh,  AAC, and MC. The evaluation metrics include WER for ASR, BLEU4 for En2Zh, METEOR, and SPIDEr for AAC, and BLEU4 and RougeL for MC.

For ASR, U-SAM achieves a WER of 2.0\% on clean speech and 5.0\% on noisy speech, outperforming most models except SALMONN, which achieves a slightly better WER of 2.1\% on clean speech and 4.9\% on noisy speech. In AST-En2Zh, U-SAM achieves a BLEU4 score of 38.3, closely approaching the best performance of 38.9 reported by MT-Rev. For AAC, U-SAM achieves the highest METEOR and SPIDEr scores (26.7\textbar{}49.0), surpassing all baselines, including GAMA (26.1\textbar{}47.8) and \cite{mei2024wavcaps} (25.0\textbar{}48.5). Similarly, in MC, U-SAM achieves a BLEU4 score of 6.0 and a RougeL score of 22.5, which are competitive with LP-MusicCaps. 

U-SAM demonstrates strong performance across all tasks, achieving superior results compared to specialized models and audio language models (ALMs), highlighting its effectiveness in handling diverse audio understanding tasks.

\begin{table}[t]
\centering
\caption{Ablation study of U-SAM on evaluation datasets.}
\label{tab:ablation_results}
\resizebox{0.5\textwidth}{!}{%
\begin{tabular}{@{}lcccc@{}}
\toprule
      & \textbf{ASR}$\downarrow$ & \textbf{En2Zh}$\uparrow$ & \textbf{AAC}$\uparrow$ & \textbf{MC}$\uparrow$ 
\\ 
\textbf{Models}                  & \%WER       & BLEU4         & METEOR\textbar{}SPIDEr       & BLEU4\textbar{}RougeL      
\\ \midrule
\textbf{U-SAM}             & (2.0, 5.0)           & 38.3                    & 26.7\textbar{}49.5                    & 6.0\textbar{}22.5                   \\ \midrule
w/o Whisper Encoder        & (39.1, 45.0)         & 14.7                    & 27.2\textbar{}50.1                    & 6.8\textbar{}23.8                   \\
w/o CED Encoder            & (2.2, 5.1)           & 39.0                    & 10.9\textbar{}24.0                    & 5.7\textbar{}20.0                   \\
w/o MERT Encoder           & (2.1, 5.2)           & 38.7                    & 26.5\textbar{}49.3                    & 2.7\textbar{}11.4                   \\
w/o TAPM                   & (5.1, 7.9)           & 31.0                    & 28.0\textbar{}51.5                    & 4.0\textbar{}16.8                   \\
w/o SACLM                  & (3.0, 6.5)           & 32.1                    & 25.0\textbar{}47.9                    & 5.2\textbar{}17.0                   \\
w/o TAPM \& SACLM         & (7.0, 8.5)           & 29.0                    & 22.7\textbar{}36.0                    & 3.5\textbar{}15.2                   \\ \bottomrule
\end{tabular}%
}
\end{table}

\subsection{Ablation Study}
To evaluate the effectiveness of the key components in U-SAM, we conduct a detailed ablation study, with results presented in Table~\ref{tab:ablation_results}. Each component is removed individually or in combination to examine its contribution across four representative tasks: ASR, En2Zh, AAC, and MC. 


The multi-encoder architecture forms the backbone of U-SAM’s unified audio modeling. It comprises three specialized encoders—Whisper, CED, and MERT—targeting speech, general audio, and music, respectively. Each encoder contributes domain-specific inductive biases that are essential for capturing diverse audio semantics. Removing the Whisper encoder results in drastic performance drops in ASR and AST, underscoring its specialization for speech understanding. Excluding the CED encoder leads to a substantial decline in AAC performance, indicating its role in modeling high-level acoustic events semantics. Similarly, removing the MERT encoder significantly impairs MC results, emphasizing its importance for extracting music-specific representations.

The TAPM facilitates task-specific feature adaptation through text prompts, enabling U-SAM to generalize effectively across tasks. Without TAPM, performance declines are observed across most tasks, particularly in AST and MC, as the model struggles to adapt to task-specific requirements. Notably, AAC performance slightly improves (METEOR 26.7 to 28.0)  due to overfitting to the dominant AAC data in the training set, further underscoring TAPM's role in preventing task overfitting and balancing task-specific learning.

The SACLM ensures refined alignment between audio and text features via contrastive learning. Removing SACLM results in performance degradation across all tasks, demonstrating its critical role in enhancing cross-modal representation.

When both TAPM and SACLM are removed, the model performance further deteriorates across all metrics, highlighting their contributions. TAPM enables task-aware adaptation, while SACLM ensures refined cross-modal alignment, together enhancing U-SAM's multi-task ability.

\section{Conclusion}
\label{sec:con}
In this work, we present U-SAM, a unified audio language model that integrates a pre-trained LLM with specialized audio encoders for speech, music, and general audio. By leveraging LoRA for efficient fine-tuning and incorporating TAPM and SACLM, U-SAM achieves robust audio-text alignment. Trained on a curated dataset, U-SAM demonstrates strong performance across various tasks and exhibits emergent capabilities on unseen tasks, underscoring its adaptability in audio language understanding and its potential for advancing multi-modal reasoning and generation.

\bibliographystyle{IEEEtran}
\bibliography{mybib}

\end{document}